\documentclass[fleqn,usenatbib]{mnras}
\usepackage{newtxtext,newtxmath}
\usepackage[T1]{fontenc}
\usepackage{ae,aecompl}

\usepackage{graphicx}	% Including Figure files
\usepackage{amsmath}	% Advanced maths commands
\usepackage{amssymb}	% Extra maths symbols
\usepackage[labelfont=bf]{caption}
\usepackage[utf8]{inputenc}
\usepackage[utf8]{inputenc}
\newcommand{\HI}{\ion{H}{i}}

\newcommand{\hkpc}{h^{-1}{\rm kpc}}
\newcommand{\hmpc}{h^{-1}{\rm Mpc}}

\newcommand{\kms}{\;{\rm km}\,{\rm s}^{-1}}

\newcommand{\gizmo}{{\sc Gizmo}}
\newcommand{\mufasa}{{\sc Mufasa}}
\newcommand{\simba}{{\sc Simba}}

\newcommand{\fedd}{f_{\rm Edd}}
\newcommand{\mbh}{M_{\rm BH}}
\newcommand{\mdot}{\dot{M}_{\rm BH}}
\newcommand{\mHI}{M_{\rm HI}}
\newcommand{\fHI}{f_{\rm HI}}

\newcommand{\RomanNumeralCaps}[1]
    {\MakeUppercase{\romannumeral #1}}

\title{Black Hole -- Galaxy Correlations in \simba}

\author[Thomas et al.]{
Nicole Thomas$^{1,2}$,\thanks{E-mail: thomas.nicolelynn@gmail.com}
Romeel Dav\'e$^{3,1,2}$,
Daniel Angl\'es-Alc\'azar$^{4}$,
Matt Jarvis$^{5,1}$
\newauthor
\\
$^{1}$Department of Physics and Astronomy, University of the Western Cape, Bellville, 7535, South Africa\\
$^{2}$South African Radio Astronomical Observatory, Observatory, 7925, South Africa\\
$^{3}$Institute for Astronomy, Royal Observatory, University of Edinburgh, Edinburgh, EH9 3HK, UK\\
$^{4}$Center for Computational Astrophysics, Flatiron Institute, 162 Fifth Avenue, New York, NY 10010, USA\\
$^{5}$University of Oxford, Denys Wilkinson Building, Keble Road, Oxford, OX1 3RH, UK}

\date{Accepted XXX. Received YYY; in original form ZZZ}

\pubyear{2019}

\begin{document}
\label{firstpage}
\pagerange{\pageref{firstpage}--\pageref{lastpage}}
\maketitle

% Abstract of the paper
\begin{abstract}
We examine the co-evolution of galaxies and supermassive black holes in the \simba\ cosmological hydrodynamic simulation.  \simba\ grows black holes via gravitational torque-limited accretion from cold gas and Bondi accretion from hot gas, while feedback from black holes is modeled in radiative and jet modes depending on the Eddington ratio ($\fedd$). \simba\ shows generally good agreement with local studies of black hole properties, such as the black hole mass--stellar velocity dispersion ($\mbh-\sigma$) relation, the black hole accretion rate vs. star formation rate (BHAR--SFR), and the black hole mass function.  $\mbh-\sigma$ evolves such that galaxies at a given $\mbh$ have higher $\sigma$ at higher redshift, consistent with no evolution in $\mbh-M_\star$.  For $\mbh\la 10^8 M_\odot$, $\fedd$ is anti-correlated with $\mbh$ since the BHAR is approximately independent of $\mbh$, while at higher masses $\fedd-\mbh$ flattens and has a larger scatter.
BHAR vs. SFR is invariant with redshift, but $\fedd$ drops steadily with time at a given $\mbh$, such that all but the most massive black holes are accreting in a radiatively efficient mode at $z\ga 2$.
The black hole mass function amplitude decreases with redshift and is locally dominated by quiescent galaxies for $\mbh>10^{8}M_{\odot}$, but for $z\ga 1$ star forming galaxies dominate at all $\mbh$. The $z=0$ $\fedd$ distribution is roughly lognormal with a peak at $\fedd\la 0.01$ as observed, shifting to higher $\fedd$ at higher redshifts.
Finally, we study the dependence of black hole properties with \HI\ content and find that the correlation between gas content and star formation rate is modulated by black hole properties, such that higher SFR galaxies at a given gas content have smaller black holes with higher $\fedd$. 
\end{abstract}

% Select between one and six entries from the list of approved keywords.
% Don't make up new ones.

%%%%%%%%%%%%%%%%%%%%%%%%%%%%%%%%%%%%%%%%%%%%%%%%%%

%%%%%%%%%%%%%%%%% BODY OF PAPER %%%%%%%%%%%%%%%%%%

\section{Introduction}
%%%%%%%%OBSERVATIONAL IDENTIFICATION OF BHS%%%%%%%%%%%%%%%%%%%%%%%
During periods of strong activity, accreting supermassive black holes (SMBHs) can be optically identified at the nuclei of their host galaxies, and are referred to as active galactic nuclei (AGN). 
Optical spectroscopic surveys are able to detect a large number of AGN, albeit being biased toward the brighter and less obscured AGN. X-rays are considered the most reliable method of AGN identification, particularly hard X-rays, owing to a significant reduction in obscuration~\citep{RG2016}.  Upcoming radio surveys such as MIGHTEE (The MeerKAT International GHz Tiered Extragalactic Exploration Survey; \citealt{Jarvis2017}) will likewise greatly assist in identifying AGN regardless of obscuration out to high redshifts.  Complementary multi-wavelength data enables characterisation of key host galaxy properties such as stellar masses and star formation rates.  By examining the relationship between AGN and galaxy properties across cosmic time, we can constrain models and assemble a more comprehensive picture of the co-evolution of galaxies and their central black holes.    

%%%%%%%%GALAXY-BH SCALING RELATIONS%%%%%%%%%%%%%%%%%%%%%%%%%%%%
It has long been recognised that
SMBH properties correlate with larger-scale properties of their host galaxies. Relations between the black hole mass $\mbh$ and properties such as the galaxy stellar mass $M_\star$ \citep{KormendyAndHo}, bulge mass $M_{\rm bulge}$  \citep{Haring2004}, bulge luminosity $L_{\rm bulge}$ \citep{Shankar2004}, velocity dispersion $\sigma$ \citep{ferrarese2000}, s\'{e}rsic index $\eta$ \citep{Grahametal2007}, and core size of massive ellipticals \citep{Thomas2016} all show fairly tight relations. There is also substantial evidence that the cosmic history of star formation rate and the evolution of global black hole accretion rate evolve quite similarly~\citep{Aird2010,Rodighiero2010,Rodighiero2015,DAA2015}
showing that galaxies and black holes grow together, albeit with large scatter in individual cases. 

%%%%%%%%%%%%%%%%%LITERATURE REVIEW%%%%%%%%%%%%%%%%%%%%%%%%%%%%
A more detailed examination highlights various ways in which galaxies are impacted by their SMBHs.  \citet{KormendyAndHo} showed that $z\approx 0$ SMBHs correlate tightly with only classical bulges and elliptical galaxies, not with disk properties, suggesting that the processes creating bulges such as mergers may also drive SMBH growth. However, this connection is much less clear at $z\sim 2-4$ during what is sometimes called the ``quasar era".  At that time, the largest SMBHs grow via radiatively efficient accretion, in contrast to locally ($z\sim 0$) where the most efficient SMBH growth happens in lower mass systems.  \citet{HeckmanAndBest} reviewed the general population of black hole-galaxy correlations to quantify where the local population of black holes reside. They divide AGN into two populations, ``jet" and ``radiative" mode AGN, corresponding to low- and high accretion efficiencies, respectively. They find that jet-mode AGN (roughly, AGN with Eddington ratios $\fedd\la 0.01$) are generally large black holes that reside in massive quenched early-type galaxies, whereas radiative mode AGN arise from moderate sized black holes in large galaxies that often still have some disk component and a pseudobulge (Type 2 Radio Quiet AGN), and that have significant star formation.
They also find that the growth of SMBHs occurs predominantly in moderately massive galaxies ($M_\star\sim 10^{10}-10^{11} M_{\odot}$) with young stellar populations.

%%%%%%%%%%%% BHs in MODELS and SIMULATIONS%%%%%%%%%%%%%%%%%%%%

The importance of black holes in galaxy formation models has grown substantially in recent years owing to the realisation that they are the most likely drivers for galaxy transformation from star-forming spirals to quenched ellipticals~\citep[e.g.][]{SomervilleDave2015}.  Models that aim to understand global galaxy--black hole co-evolution must first be shown to reproduce observed galaxy population trends within a cosmological context, in order to be physically plausible.  This has long been a major challenge for galaxy formation models.  

Pioneering work by \citet{Springel2005b} and \citet{DiMatteo2005} incorporated black hole growth and associated feedback into hydrodynamical galaxy formation simulations.  Their simulations were able to efficiently grow black holes during galaxy mergers, and the energetic AGN feedback injected thermally was able to clear out the gas and leave a quenched galaxy.  By extrapolating results from a library of merger simulations, \citet{Hopkins2006} was able to show that the resulting properties matched a number of key galaxy--black hole observables.  The resulting scenario highlighted the importance of mergers in driving rapid growth of both the bulge and black hole, culminating in a final short-lived quasar phase that blows out the remaining gas~\citep{Hopkins2007}.  \citet{DiMatteo2008} incorporated this model into cosmological runs, which was broadly successful at reproducing galaxy--black hole correlations.  However, the feedback from the black holes was ultimately unable to sufficiently quench star formation to the extent required to match observed massive galaxies.

Semi-analytic models (SAMs) of galaxy formation also included black holes and associated feedback in order to quench galaxies. In contrast to the merger-driven scenario, in SAMs quenching was found to be effective when enacted via heating of halo gas~\citep{Croton2006,Bower2006}, called ``radio mode" or ``maintenance mode" feedback~\citep{Somerville2008}. It was found that mergers were not sufficiently frequent and did not release sufficient energy to keep galaxies fully quenched, which was later confirmed using cosmological hydrodynamic simulations~\citep{Gabor2012,Gabor2015}. Thus the dominant mode for how black holes grow, and how they impact their host galaxies in order to reproduce observed scaling relations, remained controversial.

Over the last five years there has been substantial work in improving models for black hole feedback in hydrodynamic simulations.  The key observables that simulations aimed to reproduce are the numbers of quenched galaxies and the exponential cutoff in the stellar mass function, while simultaneously reproducing black hole--galaxy scaling relations. The Illustris simulation~\citep{Vogelsberger2014,Genel2014} was able to produce some quenched galaxies~\citep{Sijacki2015} by turning up the feedback strength relative to \citet{DiMatteo2008}, but was not able to reproduce the correct colour distribution or produce an exponential cutoff, and moreover too strongly evacuated massive halos of their gas~\citep{Genel2014}.  The EAGLE simulation~\citep{Schaye2015} employed quite a different feedback model, and was able to reproduce both a bimodal colour distribution~\citep{Trayford2017} and a reasonable mass function.  Illustris-TNG~\citep{Springel2018} improved upon Illustris's feedback model by including two-mode feedback~\citep{Weinberger2018} reminiscent of observations~\citep{BestAndHeckman2012}, yielding galaxy properties similar to those observed.  All these simulations, however, assumed that the black hole feedback was spherical or quasi-spherical; in contrast, the Horizon-AGN simulation~\citep{Dubois2012,Dubois2014,volonteri2016} used bipolar feedback that is more reminiscent of observed AGN feedback, but were unable to produce sufficiently quenched massive galaxies.

All the above simulations modeled black hole growth via Bondi-Hoyle-Lyttleton accretion \citep{Hoyle1939,BondiAndHoyle,Bondi1952} or variations thereof \citep{Dubois2012,Choi2012,RG2015}. Owing to Bondi accretion's squared dependence on the mass of the SMBH, black hole growth must be self-regulated by feedback from the black hole itself in order to avoid runaway growth \citep{DAA2015}.  This self-regulation is difficult to achieve without quasi-spherical feedback, yet observations of black hole feedback from jets or outflows generally show bipolar feedback.  It is worth noting that even feedback that is implemented spherically at small scales can result in bipolar outflows on larger scales, owing to collimation by the surrounding gas.

The recent \simba\ simulation~\citep{Dave2019} employed a different model for black hole growth and feedback. \simba's model is based on the idea that torques owing to disk instabilities are responsible for dissipating angular momentum and allowing the black hole accretion disk to be fed.  \citet{HopkinsQuataert2011} found that non-axisymmetric perturbations in the stellar gravitational potential produce orbit crossings and shocks that efficiently remove angular momentum even at scales $\lesssim 10$pc, and derived analytic equations for the loss of angular momentum in the presence of such shocks, which resulted in a model capable of reproducing gas inflow rates down to $\lesssim0.1$pc that matched very high-resolution simulations (orders of magnitude better than the Bondi parameterisation). \simba\ implements this gravitational torque limited model in a sub-grid manner~\citep{DAA2013,DAA2015,DAA2017,DAA2017c},
for accretion from cold gas, while still using Bondi accretion from hot gas where the Bondi assumption of gravitational capture from a hot medium is more appropriate.  \citet{DAA2013} showed that this so-called torque-limited accretion model results in galaxy--black hole scaling relations even without any self-regulating feedback, and \citet{DAA2017} showed that this result holds even in the presence of strong black hole feedback.  
Because \simba's accretion model does not require self-regulation, it is possible to implement black hole feedback in a more realistic way.
In particular, \simba\ uses bipolar outflows whose velocity increases rapidly as $\fedd$ drops, broadly motivated by the two-mode feedback seen observationally~\citep{HeckmanAndBest}. In \citet{Dave2019}, we showed that this model produces a correlation between $\mbh$ and $M_\star$ that is in good agreement with observations, along with the correct fraction of quenched galaxies as a function of stellar mass.

In this paper, we extend the preliminary results in \citet{Dave2019} to more comprehensively consider a wider range of black hole properties, their relationship to host galaxy properties, and their evolution.  We focus primarily on the predictions of black hole masses and accretion rates predicted and their correlation with galaxy properties such as $M_\star$, SFR, and \HI\ mass in \simba.  We defer a more careful comparison in the observational plane to future work. We show that in most cases \simba\ produces good agreement with available observables, and makes interesting predictions for the relationship to host galaxy properties that can be tested in future multi-wavelength surveys.

In \S\ref{sec:methods} we describe the \simba\ simulations and accretion and feedback models, 
followed by \S\ref{sec:results} which shows the resulting scaling relations in the local universe and their evolution with redshift.
We then discuss our findings and conclusions in \S\ref{sec:discussion}.

\section{Simulations}
\label{sec:methods}

We use the \simba\ simulation~\citep{Dave2019}, run with a version of the cosmological gravity+hydrodynamics code \gizmo\footnote{www.tapir.caltech.edu/~phopkins/Site/GIZMO.html} \citep{Hopkins2015} in its Meshless Finite Mesh (MFM) hydrodynamics solver. \simba\  models a $(100h^{-1}{\rm Mpc})^{3}$ comoving randomly-selected volume down to $z=0$ with $1024^{3}$ dark matter particles and $1024^{3}$ gas elements.
\simba\ includes radiative cooling and photoionisation heating using the {\sc Grackle-3.1} package~\citep{Smith2017}, assuming a \citet{HaardtMadau2012} ionising background that incorporates self-shielding on the fly via the prescription in \citet{Rahmati2013}.
Star formation is modeled by a \citet{Schmidt1959} law on the molecular hydrogen component, where the $H_2$ fraction is computed via a subgrid prescription following \citet{KrumholzGnedin2011}.  Chemical enrichment is followed for 9 metals, from Type II and Type Ia supernovae (SNe) and Asymptotic Giant Branch (AGB) stars. The Type~II SN energy is assumed to (instantaneously) drive galactic outflows, implemented via decoupled, kinetic, two-phase winds, with scalings of mass outflow rates with galaxy stellar mass based on the particle tracking results of \citet{DAA2017b} using the Feedback in Realistic Environments (FIRE) simulations~\citep{Hopkins2014,Hopkins2018}.  Energy from Type~Ia and AGB stars is also added at later times by tracking stellar evolution based on the \citet{Bruzual2003} stellar population synthesis model.

We adopt a standard $\Lambda$CDM cosmology with parameters $\Omega_{\Lambda}=0.7$, $\Omega_{m}=0.3$, $\Omega_{b}=0.048$, $h=0.68$, $\sigma_{8}=0.82$, and $n_{s}=0.97$~\citep{Planck2016}.
Cosmological initial conditions are generated using {\sc Music} \citep{Hahn2011} with the minimum comoving softening length set to $0.5\%$ of the mean interparticle distance for dark matter particles, corresponding to a full minimum softening radius of $\epsilon=1.4h^{-1}{\rm kpc}$ with a 64-neighbour cubic spline kernel. The minimum gas smoothing length is half the minimum softening length.  Further modeling details are available in \citet{Dave2019}.  Given the centrality of the black hole model to the results of this work, we present the \simba\ black hole growth and feedback models in more detail in the following sections.

\subsection{Black Hole Seeds}

Many uncertainties still remain with regards to black hole seeding \citep{Volonteri2010}. For simplicity, we do not attempt to mimic the physics of any seed formation mechanism in detail, and instead assume that a black hole appears at the center of each galaxy once it exceeds a mass where efficient black hole growth can occur, which we take to be $M_\star>10^{9.5}M_\odot$. Below this mass, higher resolution simulations have shown local stellar feedback disrupts black hole accretion and suppresses growth~\citep[e.g.][]{Dubois:2015,RG2016,DAA2017c,Habouzit2017}.
An on-the-fly fast friends-of-friends (FoF) algorithm is used to identify galaxies. If the FoF galaxy does not already include a black hole particle and is above the threshold stellar mass, we insert a seed of mass \textcolor{black}{ $M_{\rm seed}=10^{4}h^{-1}M_{\odot}$ } at the location of the most bound gas particle.  This places the black hole well below the observed $\mbh-M_\star$ relation, but as discussed in \citet{DAA2013} and also shown below, the black hole grows fairly rapidly onto the relation via torque-limited accretion.

\subsection{Black Hole Accretion}

We employ a two-mode accretion model for the growth of black holes in \simba. The first mode follows the torque-limited accretion model presented by \citet{DAA2017} for cold gas ($T<10^{5}K$), while the second mode models Bondi accretion solely from hot gas ($T>10^{5}K$).

\subsubsection{Gravitational Torque-Limited Model}
\label{sec:gt}
Accretion rates are based on the gravitational torque model of \citet{HopkinsQuataert2011} which estimates the gas inflow rate, $\dot{M}_{\rm Torque}$, driven by gravitational instabilities from galactic scales down to the accretion disk surrounding the black hole as
\begin{equation}
\label{eq:GT1}
\dot{M}_{\rm BH}=(1-\eta)\dot{M}_{\rm Torque}
\end{equation}
where $\eta=0.1$ is the radiative efficiency \citep{Marconi2004} 
and $\dot{M}_{\rm Torque}$ is estimated from properties of the host galaxy within a distance $R_{0}$ of the black hole by \textcolor{black}{
\begin{equation}
\label{eq:GT2}
\begin{split}
\dot{M}_{\rm Torque}\approx \epsilon_{T}f^{5/2}_{d} \times \Big(\frac{M_{\rm BH}}{10^{8} \rm M_{\odot}}\Big)^{1/6}\Big(\frac{M_{\rm gas}(R_{0})+M_{*}(R_{0})}{10^{9} \rm M_{\odot}}\Big)\\
\times \Big(\frac{R_{0}}{100 \rm pc}\Big)^{-3/2}\Big(1+\frac{f_{0}}{f_{\rm gas}}\Big)^{-1} \rm M_{\odot}yr^{-1}
\end{split}
\end{equation}
where $f_{d}$ is the disk fraction for the combined gas and stellar disk mass $M_{d}(R_{0})$ within $R_{0}$ such that:
\begin{equation}
\label{eq:GT3}
f_{d}\equiv \frac{M_{d}(R_{0})}{M_{\rm gas}(R_{0})+M_{*}(R_{0})}
\end{equation}
while $M_{\rm gas}(R_{0})$ and $M_{*}(R_{0})$ represent the total gas and stellar masses within $R_{0}$. $f_{gas}$ is the gas fraction relative to the disk mass
\begin{equation}
\label{eq:GT4}
f_{\rm gas}\equiv \frac{M_{\rm gas}(R_{0})}{M_{d}(R_{0})}
\end{equation}
} and 
\begin{equation}
\label{eq:GT5}
f_{0}\approx 0.31 f^{2}_{d}\Big(\frac{M_{d}(R_{0})}{10^9 \rm M_{\odot}}\Big)^{-1/3}.
\end{equation}
\noindent
We define $\epsilon_{\rm T} \equiv \epsilon_{\rm m} \times \alpha_{\rm T}$, where $\alpha_{\rm T} = 5$ is the normalization of $\dot{M}_{\rm Torque}$ proposed by \citet{HopkinsQuataert2011} and $\epsilon_{\rm m}$
is a free parameter introduced to account for processes that affect the radial transport of gas at unresolved scales. 
We tune this to $\epsilon_{m}=0.1$ in order to match the amplitude of the $\mbh-M_\star$ relation at \textcolor{black}{$z=0$}. 
$R_{0}$ is taken to be the radius enclosing 256 gas elements, with an upper limit of 2$\hkpc$. Evaluating equation \ref{eq:GT2} requires the separation of spheroidal and disk components within $R_0$, which is done by means of a kinematic decomposition \citep{DAA2013,DAA2015}.

\subsubsection{Bondi-Hoyle-Lyttleton Parameterisation}
\label{sec:bondi}
The Bondi model has been widely used as a prescription for black hole growth in galaxy formation simulations \citep[e.g.][]{Springel2005b,Dubois2012,Choi2012}. For a black hole mass $\mbh$ moving at a velocity $v$ relative to a uniform distribution of gas with density $\rho$ and sound speed $c_{s}$, the Bondi rate is given by
\begin{equation}
\dot{M}_{\rm Bondi}= \alpha \frac{4\pi G^{2}\mbh^{2}\rho}{(c_{s}+v)^{3/2}}
\end{equation}
where $\alpha$ is a dimensionless parameter usually used to boost accretion rates and partially compensate for high mean gas temperatures as a consequence of the multi-phase subgrid model of star formation and/or the lack of resolution required to resolve the Bondi radius. We do not use a boost factor and rather suppress $\dot{M}_{\rm Bondi}$ by the same factor as $\dot{M}_{\rm Torque}$ for consistency ($\alpha \equiv \epsilon_{\rm m}=0.1$).

\subsubsection{Numerical Implementation}

We apply the torque-limited accretion formula to all the gas within $R_0$ that has a temperature $T<10^5$K, while for $T>10^5$K gas we employ the Bondi formula, computing $\rho$ and $c_s$ from the hot gas only within $R_0$. A given black hole can thus accrete gas in both Bondi and torque-limited modes at any given timestep.  The total accretion onto the black hole is then
\begin{equation}
\label{eq:totaccr}
\dot{M}_{\rm BH}=(1-\eta)(\dot{M}_{\rm Bondi}+\dot{M}_{\rm Torque})
\end{equation}
We limit Bondi accretion to the Eddington rate, while torque-limited accretion is capped at $3\times$ the Eddington rate. Black holes are further limited to not grow beyond $0.1\%$ of their mass in a single simulation time step to avoid large stochastic fluctuations, but this limit is very rarely invoked.

Black hole accretion proceeds stochastically \citep{Springel2005}. Gas particles within $R_{0}$ get a fraction of their mass subtracted and added to the black hole, with a probability that statistically satisfies the mass growth (eq. \ref{eq:totaccr}).  If a particle is sufficiently small compared to its original mass, it is swallowed completely.

\subsection{Black Hole Feedback}

\simba\ employs three black hole feedback mechanisms: (a) Radiative feedback in high-$\fedd$ black holes; (b) Jet feedback in low-$\fedd$ black holes; (c) X-ray feedback. The first two are implemented purely kinetically and purely bipolar, with the direction set as $\pm$ the direction of the angular momentum of gas within $R_0$.  Albeit that \citet{Whittam2018} reports a more continuous distribution of $\fedd$ if the radio AGN sample are selected from deeper radio observations, the motivation for this is based on the observed dichotomy of black hole accretion rates and the corresponding properties of their outflows \citep{HeckmanAndBest}. At high $\fedd$ ($\ga$few percent), AGN are observed to drive multi-phase winds at velocities of $\sim10^{3} \kms$ that include warm molecular and ionised gas. At low Eddington ratios, AGN mostly drive hot gas in collimated jets at velocities $\sim 10^{4} \kms$, that can be seen to inflate super-virial temperature bubbles in the surrounding hot gas. Within jet modes, this dichotomy can be seen between ``high excitation" (HERG) and ``low excitation" radio galaxies (LERG). The former are found typically in lower mass, bluer host galaxies, and the latter in more massive, earlier types. \simba\ thus models AGN feedback in such a way as to directly mimic the energy injection into large scale surrounding gas using bipolar feedback with properties taken as much as possible from AGN outflow observations. 

\subsubsection{Kinetic Feedback}
For high $\fedd$ mode outflows, an outflow velocity is chosen based on ionised gas linewidth observations of X-ray detected AGN from SDSS \citep{Perna2017a} and parameterised in terms of the black hole mass such that
\begin{equation}
    v_{w,EL} = 500+500(\log\ M_{\rm BH}-6)/3 \kms 
\end{equation}
and are referred to as AGN winds.\\
If $\fedd<0.2$, we slowly transition to the jet mode where the velocity becomes increasingly higher as $\fedd$ drops:
\begin{equation}
    v_{w} = v_{w,EL}+7000\log\Big(\frac{0.2}{f_{\rm Edd}} \Big) \kms,
\end{equation}
with a cap to the velocity increase at 7000$\kms$. Additionally, a criterion requiring $\mbh>M_{\rm BH,lim}$ is added and motivated by observations that show that jets only arise in galaxies with velocity dispersions corresponding to black holes with $\mbh\gtrsim 10^{8} \rm M_{\odot}$~\citep{Mclure2004,Barisic2017}. We conservatively choose $M_{\rm BH,lim}=10^{7.5} \rm M_{\odot}$. This mass limit prevents small black holes with temporary low accretion rates from driving high powered jets.

AGN-driven outflows are modelled by stochastically kicking particles around the black holes with velocity $v_w$ with probability 
\begin{equation}
\label{eq:FB1}
p_{j} = \frac{1-f_{m}}{f_{m}} \times \frac{w_{j}}{m_{j}} \times \dot{M}_{\rm BH} \Delta t
\end{equation}
where $w_{j}$ is a kernel weight and $f_{m}$ is the fraction of mass accreted by the black hole and subtracted from the gas particle before ejection. This gives an outflow mass loading factor of $\dot{M}_{out}/\dot{M}_{\rm BH} = (1-f_{m})/f_{m}$.  We set $f_m$ for each outflow event such that the momentum ejected by the black hole is $20L/c$, where $L=\eta \dot{M}_{\rm BH} c^2$.

\subsubsection{X-ray Feedback}
The energy input rate due to X-rays emitted by the accretion disk is computed according to \citet{Choi2012}, assuming a radiative efficiency $\eta=0.1$.  In gas-rich galaxies, severe radiative losses are expected in the ISM, hence we only apply X-ray feedback below a galaxy gas fraction threshold of \textcolor{black}{$f_{\rm gas}<0.2$}, and in galaxies with full velocity jets ($v_w\ga 7000\kms$).  For further details see \citet{Dave2019}.

\subsection{Analysis of Simulations}

\simba\ outputs 151 snapshots to $z=0$, but here we will be primarily concerned with snapshots at $z=0,0.5,1,2,3$.  Galaxies are identified as gravitationally bound collections of gas and star particles using a friends-of-friends galaxy-finder. Black holes are assigned to the galaxy to which they are most gravitationally bound and thus galaxies may have many black holes. We consider the largest black hole within the galaxy to be the central black hole and refer to this as the black hole mass. Typically, the other black holes are much smaller and add no significant mass relative to the central black hole. Galaxies are post-processed using the {\sc YT}-based package {\sc caesar}\footnote{caesar.readthedocs.io/} which outputs all pre-computed galaxy information and key properties in a convenient hdf5 catalogue.  All results here are obtained from {\sc caesar} catalogues which are generated from simulation snapshots at specified redshifts.

More details about the simulations and black hole accretion and feedback models can be found in \citet{DAA2013,DAA2015,DAA2017,Dave2019}. 

\section{Results}
\label{sec:results}

Black hole properties are observed to be correlated with properties of their host galaxy. In this section we show predictions for key black hole--galaxy scaling relations, as well as distributions of black hole properties. The goal is to assess how well \simba\ reproduces the observed supermassive black hole population, and to identify trends in black hole properties arising from the input physics.

In \simba, black hole particles carry two properties: the black hole mass $\mbh$, and the instantaneous black hole accretion rate $\mdot$ (eq.~\ref{eq:totaccr}).  From these, it is possible to compute the Eddington ratio 
\begin{equation}
\label{eq:fedd}
    \fedd=\frac{\eta c \sigma_T}{4\pi G m_p}\frac{\mdot}{\mbh}
\end{equation}
In this paper, we will focus on relating these intrinsic black hole properties to the global galaxy properties such as stellar mass $M_\star$, stellar velocity dispersion $\sigma$, star formation rate (SFR), and \HI\ fraction $\fHI = \mHI/M_\star$. We defer a comparison in terms of observational quantities such as AGN luminosities in various bands to future work.

\subsection{Black hole growth history}
\label{sec:bhgrowth}

Black holes and galaxies appear to grow at commensurate rates when viewed globally over cosmic time~\citep{Aird2010,MadauDickinson2014}, even though there is only a weak correlation between the instantaneous growth rates for individual systems~\citep[e.g.][]{Hickox2014}.  In this section we examine the global growth of galaxies relative to black holes over cosmic time, in order to situate the forthcoming discussion of the relationship between black holes and their host galaxies.

\begin{figure}
\begin{center}
\includegraphics[width=0.9\linewidth]{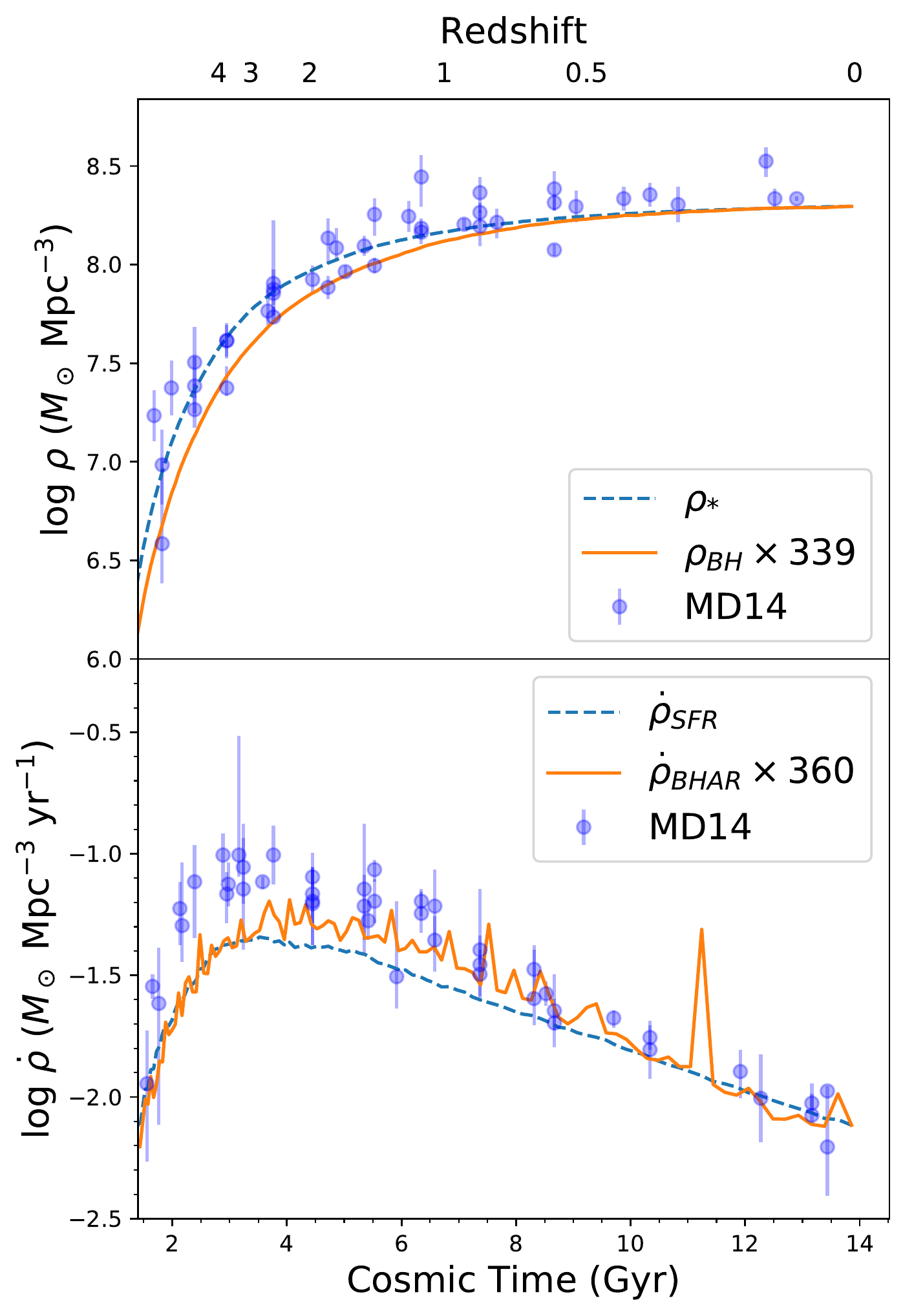}
\captionof{figure}{{\textit Top:} The evolution of the mass density in stars (blue dashed) and black holes (orange) in \simba, with the black hole mass density multiplied by 339 to match the values at $z=0$.   {\it Bottom:} Analogous plot showing the global star formation rate density (blue dashed) and black hole accretion rate density (orange), with the latter multiplied by 360.  In both panels, observations are shown from the compilations of stellar mass and SFR densities by \citet{MadauDickinson2014}. Globally-averaged stellar and black hole growth track each other well over time.}
\label{fig:bhevol}
\end{center}
\end{figure}

Figure~\ref{fig:bhevol}, top panel, shows the global mass density in black holes vs. that in stars in \simba's resolved galaxy population over cosmic time.  The bottom panel shows the global SFR density vs. the BHAR density.  In each case, the galaxy quantity is shown as the blue dashed line, and the black hole quantity is shown as the orange solid line.  The black hole curves have been multiplied by a factor that results in these quantities being equal at $z=0$, and this factor is indicated in the legend.  Data points are shown for the {\it galaxy} quantities, namely the stellar mass density and the SFR density, from \citet{MadauDickinson2014}, so should be compared with the blue dashed curves.  Note that the bottom panel is similar to the Lilly-Madau plot shown in \citet{Dave2019}, except here we are considering only resolved galaxies; this makes a very minor difference.

In a globally averaged sense, the black holes and galaxies in \simba\ track each others' growth fairly closely.  Since black holes are only seeded once a galaxy reaches a certain mass, the mass density in black holes lags slightly behind the stellar density at early epochs. Conversely, once the black hole is seeded, it has to grow a bit more rapidly than the stars in order to catch up, as seen in the bottom panel.  Comparing to observations of the stellar mass density growth, \simba\ shows excellent agreement.  For the SFR density evolution, \simba\ falls short by a factor of two during Cosmic Noon \citep[as discussed in][]{Dave2019}.

The values of the multiplicative factors are interesting.  Overall, it is observed that typically $\mbh\approx 0.0014M_\star$~\citep{Sun2015}, which would suggest the required multiplicative factor should be about 670. Instead, it is $\approx\times 2$ lower.  This arises because there is substantial scatter in the $\mbh-M_\star$ relation (Figure~\ref{fig:mbhms}), and the over-massive black holes contribute more than their share to the black hole mass density budget.  Overall, the prediction is thus within the expected range.

In contrast, for the global instantaneous growth rate, the multiplicative factor is an order of magnitude lower in \simba\ than inferred from observations~\citep{Aird2010,MadauDickinson2014}.  This may arise because in \simba, essentially all galaxies that have black holes are ``active", in the sense that their SMBHs are accreting at a nonzero rate. However, many of these may not have sufficient accretion to be observable as AGN, and hence their accretion would not be counted observationally towards the global BHAR. Indeed, during much of cosmic time, the most rapidly growing black holes are the ones in moderate-mass star-forming systems, for which it is quite difficult to identify a black hole unless it is in a (rare) Seyfert phase.  In \simba, a substantial amount of accretion occurs in such galaxies, as we will discuss later.  Since the integral of the SFRD and BHARD should give the cosmic stellar mass and black hole mass density (modulo stellar evolution effects), the factor for mass growth and that for instantaneous growth should be comparable, which it is for \simba, but not so for observed values. This suggests that there is substantial instantaneous black hole growth missed by current surveys of AGN across cosmic time \citep{HickoxAlexander2018}. Other simulations have found similar results: Illustris \citep{Sijacki2015} find a BHARD/SFRD $\sim 1000$, while \citet{Weinberger2017} used the {\sc Arepo} code \citep{Springel2010} to model AGN and black hole growth, which produces a SFRD 0.3~dex lower than observations and BHARD/SFRD of a few hundred. 

Overall, black holes in \simba\ grow commensurately with galaxies in a globally-averaged sense, which is qualitatively consistent with observations.  A naive quantitative comparison suggests that \simba\ overproduces the global accretion rate density at all epochs, but observational selection effects may play a significant role in this; we will look at this more closely in future works. We next examine the scaling relations for individual galaxies, to identify the galaxies where black holes have grown and are still growing.

\subsection{Black Hole--Galaxy Scaling Relations}

\subsubsection{$\mbh-M_\star$ relation}
\label{sec:mbhms}

\begin{figure*}
\begin{center}
\includegraphics[width=0.9\linewidth]{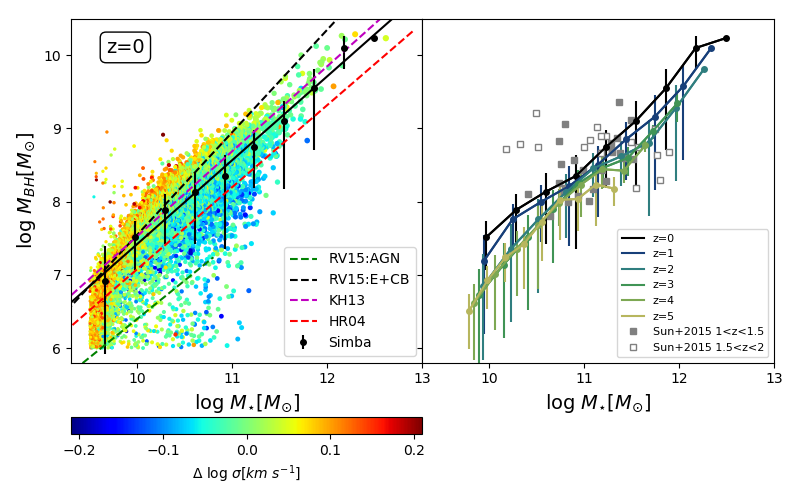}
\captionof{figure}{\textit{Left:} $\mbh-M_\star$ relation at \textcolor{black}{$z=0$}. Each circle represents a \simba\ galaxy coloured by \textcolor{black}{deviation from the $M_{\star}-\sigma$ relation at a fixed $M_{\star}$ bin}. Black points show the running median, while the solid black line shows a linear best fit to those points. The green, black, magenta, and red dashed lines show observations by \citet{Reines2015} for AGN as well as for ellipticals and classical bulges, \citet{KormendyAndHo}, and \citet{Haring2004} respectively. \textit{Right:} Evolution of the $\mbh-M_\star$. Each coloured line represents the running median at a given redshift. The filled and empty grey squares show observations from \citet{Sun2015} for $1<z<1.5$ and $1.5<z<2$ respectively. \simba\ predicts a roughly unevolving correlation between the stellar growth of a galaxy and its central SMBH which appears to be in agreement with observations shown here.}
\label{fig:mbhms}
\end{center}
\end{figure*}

A basic property of galaxies is their total stellar mass.  There is a correlation between $M_\star$ and $\mbh$ that is fairly tight and roughly linear for bulge-dominated galaxies, while disk-dominated systems show a large scatter and tend to lie below the $\mbh-M_\star$ relation~\citep{Haring2004,Mclure2006,Reines2015,KormendyAndHo,McConnell2013,Graham2016}.
Local black hole mass measurements based on stellar and gas kinematics can introduce uncertainties,
but here we will compare to derived black hole masses without considering such observational effects.

The left panel of Figure \ref{fig:mbhms} shows the $\mbh-M_{*}$ relation at $z=0$ produced by \simba.  This is as shown in \citet{Dave2019}, except here each galaxy is coloured by \textcolor{black}{the deviation of its stellar velocity dispersion $\sigma$, from the median $M_{\star}-\sigma$ relation}.  The black circles with error bars show the median $\mbh$ in a given $M_{*}$ bin with $1\sigma$ scatter around the median.  The green, black, magenta, and red dashed lines show observations from~\citet{Reines2015} for AGN as well as for ellipticals and classical bulges, \citet{KormendyAndHo}, and \citet{Haring2004} respectively. From here on, we consider only $\mbh>10^{6} M_{\rm \odot}$ and $M_{*}>10^{9.5} M_{\rm \odot}$ as this is roughly the point above which black holes are approaching the $\mbh-M_\star$ relation, and thus become broadly insensitive to the details of the seeding prescription \citep[for a full description see][Figure 13]{Dave2019}.

\simba\ produces a clear correlation between the stellar mass of a galaxy and the mass of its black hole. For a best-fit linear relation \textcolor{black}{to the median in the form} 
$\log [\mbh/M_{\odot}] = \alpha\ \log[M_{*}/10^{11}M_{\odot}] +\beta$, we find \textcolor{black}{ $\alpha = 1.147 \pm 0.002$ and $\beta=8.568 \pm 0.214$  for all galaxies.  If we divide the sample into star-forming and quenched at sSFR$=10^{-10.8}$yr$^{-1}$, we obtain for the quenched sample $(\alpha,\beta)=(1.071 \pm 0.002,8.651 \pm 0.205)$.}  This is in the range of observations of early-type galaxies, from \citet{KormendyAndHo} who find $(\alpha,\beta)=(1.16,8.69)$ to \citet{Haring2004} who find $(\alpha,\beta)=(1.12,8.20)$, as is evident from the \simba\ data mostly lying in between these two observational fits.  For the star-forming sample, the slope is poorly constrained because there is large scatter, but if we fix the slope to the observational value of $\alpha=1.05$~\citep{Reines2015}, then our amplitude \textcolor{black}{$\beta=8.231 \pm 0.824$} is somewhat higher than the value of $\beta=7.45$ obtained by \citet{Reines2015} for AGN.

\textcolor{black}{There is also a strong connection between the mass of the central black hole and the deviation of the stellar velocity dispersion of its host galaxy, $\sigma$, from the median $M_{\star}-\sigma$ relation. At a given stellar mass, larger black holes live in galaxies with higher stellar velocity dispersions. This trend is indicative of the tight correlation between $\mbh$ and $\sigma$ that we will discuss in \S\ref{sec:msigma}.}
%There is also a strong trend between the mass of the central black hole and the \HI\ fraction of its host galaxy.  At a given stellar mass, smaller black holes live in more gaseous, \HI\ rich galaxies.  As discussed in \citet{Dave2019}, black hole growth seems to be a primary determinant for quenching a galaxy, and here it commensurately seems to lower their \HI\ content as well. We will examine the black hole properties in relation to \HI\ content in more detail in \S\ref{sec:HI}.

The right panel of Figure~\ref{fig:mbhms} shows the evolution of the $\mbh-M_{*}$ relation for redshifts $z=0-5$. We compare  to observations by \citet{Sun2015} of Herschel-detected broad line AGN (BLAGN) for $1<z<2$, shown as the squares. \simba\ predicts very little evolution in the $\mbh-M_{*}$ relation, less than a factor of two over this redshift range. This is generally in agreement with the observations shown here, as well as with \citet{Shields2003} who found no distinct evolution in the $\mbh-M_{*}$ relation for quasars out to $z\sim3$, and \citet{Mullaney2012} who used X-ray stacking analyses to determine that $\mbh/M_{*}$ was approximately constant and independent of redshift. In contrast, \citet{Mclure2006} looked at massive ($M_{*}\sim10^{12}M_\odot$) early type galaxies and found a change in the ratio of $\mbh/M_{\rm sph}$ (where $M_{\rm sph}$ is the spheroidal mass component, $\approx M_\star$ for massive galaxies) increasing by $\sim\times 4$ out to $z\simeq2$.  However, AGN selection tends to bias high-$z$ samples increasingly towards over-massive black holes, which can mimic evolution~\citep{Lauer2007}.  Thus the actual amount of evolution is not precisely determined, but appears to be generally modest, as \simba\ predicts.

\subsubsection{$\mbh-\sigma$ Relation}
\label{sec:msigma}
\begin{figure*}
\begin{center}
\includegraphics[width=0.9\linewidth]{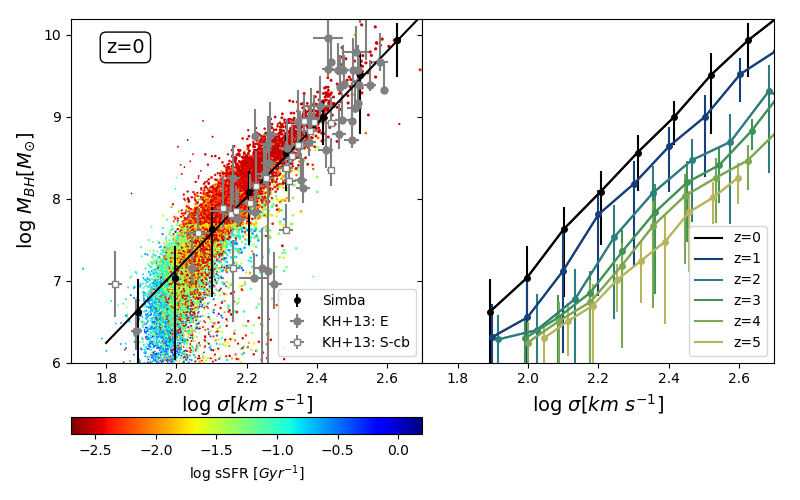}
\captionof{figure}{\textit{Left:} $M_{\rm BH}-\sigma$ relation at \textcolor{black}{$z=0$}. Each circle represents a \simba\ galaxy coloured by sSFR. Black points show the running median, while the solid black line shows a linear best fit to those points. The filled grey circles show observations of a sample of black holes measured in elliptical galaxies, and the empty grey squares are those of spiral galaxies with classical bulges compiled by \citet{KormendyAndHo}. \textit{Right:} Evolution of the $M_{\rm BH}-\sigma$ relation. Each line represents the running median at a given redshift. Results from \simba\ are in close agreement with the tight relation between the mass of a SMBH and galaxies with a significant bulge counterpart found by observations. The normalization of this correlation decreases with redshift, as expected for a universal $\mbh-M_\star$ correlation. }
\label{fig:mbhsig}
\end{center}
\end{figure*}

It has long been recognised that present-day black holes correlate more closely with properties of a galaxy's bulge mass than its total stellar mass~\citep[see][and references therein]{KormendyAndHo}. \citet{KormendyAndHo} argue that since bulges and elliptical systems formed from galaxy mergers, this argues for a connection between black hole growth and galaxy mergers.  In contrast, pseudobulges, which are more related to the secular evolution of a disk galaxy, do not satisfy the tight BH-galaxy correlations.
\citet{ferrarese2000} and \citet{Gebhardt2000} showed that the correlation with black hole mass is tightest with galaxy stellar velocity dispersion, the so-called $\mbh-\sigma$ relation.  Hence it is instructive to examine this relation in \simba\ as a test of this connection.

Figure \ref{fig:mbhsig}, left panel, shows the $z=0$ $M_{\rm BH}-\sigma$ relation for \simba\ galaxies. The 1-D velocity dispersion $\sigma$ is calculated from individual star particles that are members of each galaxy; \textcolor{black}{we have not applied an aperture correction}. 
The individual galaxy points are coloured by sSFR, with the black circles with errorbars showing the median $\mbh$ in a given $\sigma$ bin. The observed $M_{\rm BH}-\sigma$ values for ellipticals (filled grey circles) and spirals with classical bulges (empty grey squares) from \citet{KormendyAndHo} are overlaid, with errorbars. Due to the large scatter in the relation for spirals with pseudobulges, these were not plotted here. 

\simba\ produces a reasonably tight $\mbh-\sigma$ relation that is in the range of the observations for massive galaxies with $\sigma\ga 200$~km/s.  Fitting \textcolor{black}{the median sample} to $\frac{M_{\rm BH}}{M_{\odot}} =\ 10^{\alpha}\ \Big(\frac{\sigma}{200 \, {\rm km\,s^{-1}}}\Big)^{\beta} $, \textcolor{black}{
we find best-fit values of $\alpha=8.472 \pm 0.006$ and $\beta=4.45 \pm 0.032$. } Observations by
\citet{Gebhardt2000} found a somewhat shallower slope $\alpha = 8.08 \pm 0.2$ and $\beta=3.75 \pm 0.3$ which is more consistent with the simulations results from \citet{DeGraf2015} ($\alpha = 8.46$, $\beta=3.49$). Nonetheless, the simulated galaxies seem to lie in the same region as the \citet{KormendyAndHo} data, with a comparable scatter.
At low-$\sigma$, there are few observed true bulges, but the \simba\ galaxies may have somewhat over-massive black holes in this range.  We note that smaller galaxies in our simulations can have potentially inaccurate stellar velocity dispersions owing to lower particle numbers with which to compute the dispersions, which may bias the $\sigma$ values low.  Nonetheless, for bulge-dominated galaxies the predicted $\mbh-\sigma$ relation in \simba\ nicely tracks observations.
\textcolor{black}{This agreement may be surprising because \citet{Shankar2016} argued that, owing to biases in the measurements of black hole masses, the observed $\mbh-M_\star$ relation and the $\mbh-\sigma$ relation are inconsistent with each other, so it is surprising that \simba\ can match both simultaneously, which other models have had some difficulty doing~\citep{Barausse2017}.  A relevant aspect of \simba\ is that at a given $M_\star$, large black holes live in higher-$\sigma$ galaxies, as shown by the colour-coding in Figure~\ref{fig:mbhms}.  This bias qualitatively mimics that seen in observations, and implies that the $\mbh-\sigma$ relation in \simba\ is not fully described by the $\mbh-M_\star$ relation convolved with the mean $M_\star-\sigma$ relation.}

Notably, the scatter in the $\mbh-\sigma$ relation is clearly higher, typically by $\sim\times 2$, compared with the scatter in the $\mbh-M_\star$ relation (Figure~\ref{fig:mbhms}).  This is opposite to the trend generally inferred from observations~\citep{ferrarese2000}.  Since $\sigma$ responds to total mass while $M_\star$ only measures the stellar mass, there can be scatter introduced between these owing to variations in the dark matter content within the stellar region.  In observations, the tight trend with velocity dispersion as well as central ($<1$~kpc) surface density~\citep{Zolotov2015} suggests that central spheroidal growth is connected to black hole growth, indicative of structurally disruptive processes such as mergers driving both.  At face value, this seems inconsistent with the growth of black holes via torque-limited accretion that associates black hole growth primarily with disk instabilities.  However, we note that recent observations tend to paint a picture where only the brightest AGN are fueled by mergers, whereas most (Seyferts-like) AGN are not~\citep[e.g.][]{Donley2018}.  

Figure \ref{fig:mbhsig}, right panel, shows the evolution of the running median of the $\mbh-\sigma$ relation.  There is a clear evolution, in that galaxies at a given $\sigma$ have lower $\mbh$ at higher redshift.  This can be understood as a consequence of the invariance of the $\mbh-M_{*}$ relation with redshift together with size evolution, related via $\sigma\propto \sqrt{GM/R}$.  If $M_\star$ is the dominant mass component, or at least is a good tracer thereof, then this equation implies that the typical galaxy size $R$ at a given mass must decrease with redshift. Indeed, \citet{Conselice2014} describes the change in the effective radius $R_e$ for $M_{*}>10^{11}M_{\odot}$ galaxies by a power-law $R_{e}\sim(1+z)^{\beta}$ where $\beta \sim -0.82\ \rm to -1.5$ depending on whether one is observing disk-like or spheroid-like galaxies.
Such a trend is qualitatively seen in \simba, and will be detailed in a forthcoming paper~(Appleby et al., in preparation).
Other simulations have also found a similar trend. \citet{DeGraf2015} find increasing slope and decreasing normalisation of the $\mbh-\sigma$ relation along with insignificant evolution in the $\mbh-M_{*}$ relation for $z=0\to4$. \citet{Sijacki2015}, on the other hand, find a flattening of the best-fit slope and decrease in the normalisation of the $\mbh-\sigma$ relation with a $\mbh-M_{*}$ relation that increase from 1.21 to 1.28 in slope but staying roughly constant in normalisation for $z=0\to4$.

Overall, \simba\ reproduces the observed relationship between stellar velocity dispersion and black hole mass, for moderate mass galaxies and larger.  However, the scatter in this relationship is clearly larger than that in the $\mbh-M_\star$ relation, which may be in tension with observations.  We will investigate this further in future work with higher-resolution simulations that can more robustly model galaxy kinematic structure along with black hole growth.

\subsubsection{BHAR-SFR relation}

\begin{figure*}
\begin{center}
\includegraphics[width=0.9\linewidth]{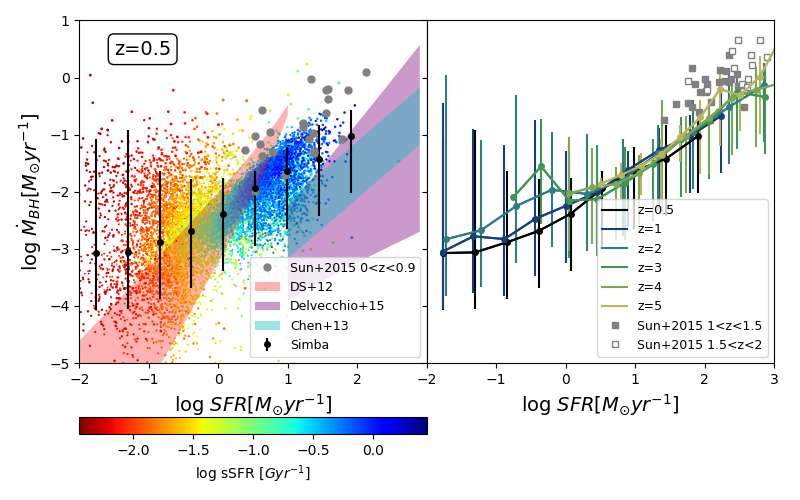}
\captionof{figure}{\textit{Left:} Black Hole Accretion Rate (BHAR or $\dot{M}_{\rm BH}$) versus Star Formation Rate (SFR) relationship at $z=0.5$. Points are colour-coded by sSFR. The black circles with error bars show the running median. The cyan and purple bands show observations of highly star forming galaxies from \citet{Chen2013} and \citet{Delvecchio2015} respectively. The red band shows observations by \citet{DS2012} and grey filled circles are observations from \citet{Sun2015}. \textit{Right:} Evolution of the $\dot{M}_{\rm BH}-SFR$ relation. Each coloured line represents the running median at a given redshift. Filled and empty squares show observations by \citet{Sun2015} for $1<z<1.5$ and $1.5<z<2$ respectively. The relation is tight for star-forming galaxies with a slope similar to that of observations and increasing scatter at low star formation rates. The $\dot{M}_{\rm BH}-SFR$ does not evolve significantly with redshift.}  
\label{fig:bharsfr}
\end{center}
\end{figure*}

Since galaxies and black holes grow commensurately in a globally-averaged sense, one expects that the SFR and BHAR should be correlated.  Indeed, such a correlation has been observed~\citep{Mullaney2012,Chen2013,Delvecchio2015}, although measuring BHARs accurately remains challenging and time variability may obscure a direct, instantaneous SFR-BHAR connection \citep{Hickox2014}.  For torque-limited accretion, \citet{DAA2015} found via post-processing of a cosmological simulation a reasonably tight connection between the SFR and nuclear activity of galaxies when averaged over galaxy dynamical timescales, although instantaneous measures could still have a large scatter.  In this section we examine the instantaneous BHAR--SFR correlation in \simba, to determine whether this connection persists in the self-consistent black hole accretion model, and how it fares versus observations.

Figure \ref{fig:bharsfr}, left panel, shows the relationship between BHAR and SFR for \simba\ galaxies, at $z=0.5$.  Each galaxy is represented by a circle colour-coded by sSFR.
The black circles with errorbars show the median BHAR in a given SFR bin. The red band shows the total SFR--BHAR relation obtained by \citet{DS2012} for Seyfert galaxies and the cyan and purple bands show results from \citet{Chen2013} and \citet{Delvecchio2015} respectively for star forming galaxies. The grey filled circles are observations of BLAGN from \citet{Sun2015} We choose to show our results at $z=0.5$ as it is comparable to the observations that we compare to, but \simba\ predicts little evolution in this relation as can be seen from the right panel.

In \simba, the SFR broadly traces the BHAR.
Note that the slope obtained by \simba\ is flatter than that of the observations.  At $z=0.5$, a fit to \textcolor{black}{the median in the form} $\log \dot{M}_{\rm BH} = \alpha \log {\rm SFR} + \beta$ for star forming galaxies yields \textcolor{black}{$\alpha=0.690 \pm 0.002 $ and $\beta = -2.34 \pm 0.003$}.  The relation clearly flattens at the lowest SFR values, owing to Bondi accretion starting to contribute significantly, which breaks the connection between star formation and black hole growth from torque-limited accretion.
The $1\sigma$ scatter around the median relation is $\approx ^{+0.5}_{-1.0}$ dex at the highest SFRs, increasing somewhat to lower SFRs.

Observations of the BHAR--SFR relation tend to focus on star-forming galaxies. The \simba\ predictions broadly lie within the region spanned by the observations. \citet{Chen2013} studied IR selected star forming galaxies at $250\mu m$ with AGN selected by X-ray and mid-IR criteria and found a slope of $\alpha=1.05$ and normalisation, $\beta=-3.72$.
\citet{DS2012} estimates the BHAR from \ion{O}{IV} ($25.89\mu m$) flux measurements and the SFR from the $11\mu m$ aromatic feature for Seyfert galaxies at a median distance of 22~Mpc, while \citet{Sun2015} uses X-ray and IR data to infer the SFR and BHAR for BLAGN. \citet{Delvecchio2015} also uses IR selected star forming galaxies and have AGN identified by X-ray criteria, similarly to the \citeauthor{Chen2013} approach.  In detail, the observations typically show lower BHAR at a given SFR in the SFR range overlapping with \simba\ predictions.  It is worth noting that the observations are generally for highly star-forming galaxies, as indicated by the shaded region for \citet{Chen2013}, except for \citet{DS2012} which samples SFRs from $\sim 0.01-10 M_\odot$yr$^{-1}$.  Given the uncertainties in determining BHARs from data at particular wavelengths, such as the difficulty in disentangling AGN emission from SF-produced IR emission or X-ray binary contributions, it remains to be seen if this discrepancy is serious.  Nonetheless, for star-forming systems, the predicted and observed slopes are quite similar.

The right panel of Figure \ref{fig:bharsfr} shows the evolution of the BHAR-SFR relation, which remains roughly unchanged from $z=5\to 0$. The main evolutionary trend is that the low-SFR tail becomes populated at lower redshifts, but in all cases it appears that there is a gradual upturn towards the lowest SFR values. This generally agrees with the \citet{Sun2015} observations represented by grey squares.  A similar lack of evolution has been found out to $z\sim 2$ by \citet{Mullaney2012}, who found a constant BHAR--SFR ratio up to $z\sim2$ which then produces the $\mbh-M_{*}$ relation found in Figure \ref{fig:mbhms}.  It is interesting that this ``AGN main sequence", as denoted by \citet{Mullaney2012}, constrained to match a roughly non-evolving $\mbh-M_\star$ relation, requires higher accretion rates than other observational determinations, and even higher than \simba\ predictions. As mentioned earlier, the integral of the BHAR and SFR should reflect the correlation between $M_\star$ and $\mbh$. \simba\ satisfies this constraint by construction, but applying it to observations may provide valuable constraints on BHAR evolution.

Overall, the BHAR--SFR relation in \simba\ shows a reasonable correlation, but with substantial scatter.  This is consistent with the idea that black holes and galaxies grow commensurately on cosmological timescales, but not necessarily on inner galactic timescales.  The BHARs in \simba\ are broadly in the range of observed values, although they appear to be somewhat higher than some recent observations; it remains to be seen whether this discrepancy constitutes a significant failing of the model.

\subsubsection{Eddington Ratios}
\label{sec:edd}

\begin{figure*}
\begin{center}
\includegraphics[width=0.75\linewidth]{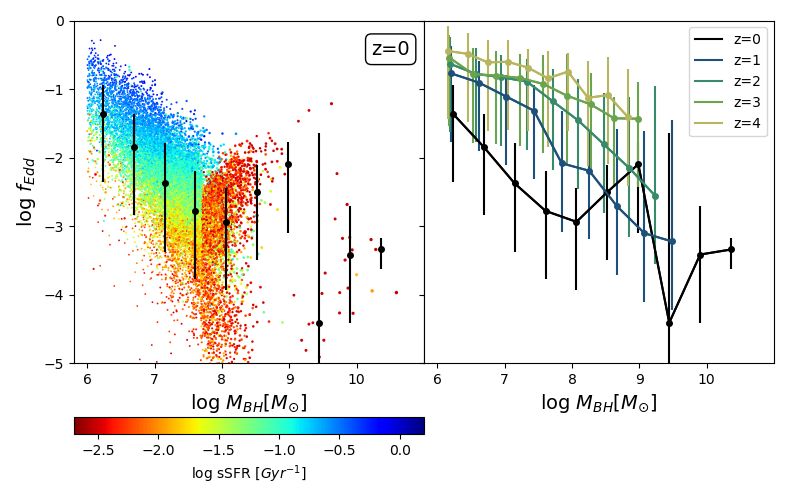}
\captionof{figure}{\textit{Left:} Eddington ratios of \simba\ galaxies as a function of the central black hole mass. Each circle represents a galaxy coloured by it's sSFR. The black circles with errorbars show the running median at $z=0$. \textit{Right:} Evolution of the Eddington ratios as a function of the central black hole mass. Each line represents the running median at a given redshift. The Eddington ratio and black hole mass show a clear anti-correlation with a slope that flattens with redshift and at which most black holes are accreting above 1$\%$ of Eddington by $z\sim2-3$. At $z=0$, the more efficiently accreting black holes tend to live in more star-forming hosts.} 
\label{fig:feddtot}
\end{center}
\end{figure*}

The Eddington ratio (eq.~\ref{eq:fedd}) appears to play a critical role in governing black hole accretion and feedback processes~\citep{HeckmanAndBest}. Observationally, AGN are often split into two broad categories, radiatively efficient with $\fedd\ga$few percent, and radiatively inefficient with $\fedd\la 0.01$.  \simba's AGN feedback model is motivated by this observed dichotomy, with $\fedd$ being the key quantity that transitions from the radiative feedback mode that has a relatively minimal impact on galaxy growth, to the jet feedback mode which plays a crucial role in quenching~\citep{Dave2019}.  Observations of $\fedd$ span a wide range, from quasars that approach values of unity and beyond~\citep{Liu2019}, to inefficiently accreting black holes in massive ellipticals that have as low as $\fedd\sim 10^{-5}$, yet are still active as evidenced by their radio jets.  Hence the Eddington ratio is an important quantity to examine.

Figure \ref{fig:feddtot}, left panel, shows $\fedd$ versus $\mbh$ for $z=0$ \simba\ galaxies colour-coded by their sSFR. The black circles with errorbars shows the median $\fedd$ value in a given black hole mass bin.  The right panel shows the running median of this relation at redshifts from $z=0-4$.

\simba\ produces a wide range of Eddington ratios, qualitatively consistent with observations.  There is a clear anti-correlation of $\fedd$ with black hole mass for $\mbh\la 10^8M_\odot$.  Above this mass, the scatter blows up, and there is a much wider range of $\fedd$.  There is a shelf at around $\mbh\sim 5\times 10^7M_\odot$ dividing these two regions, above which there is a strong increase in low-sSFR galaxies.  This is our minimum black hole mass for the onset of jet feedback, and shows that the jet feedback is directly responsible for quenching galaxies.  

For lower-mass black holes, one can see a strong dependence in the $\fedd$ on sSFR, where black holes in star forming galaxies accrete more efficiently at a given black hole mass.  This is consistent with torque-limited accretion~(eq.~\ref{eq:GT2}), in which increasing disk mass and gas fraction drive black hole accretion, accompanying an increase in star formation which is driven by these same factors (except over the entire disk, rather than the inner disk). In this regime, AGN feedback has only a minor impact~\citep{DAA2017,Dave2019}, so the growth of both stars and black holes is supply-limited.  

At high black hole masses, $\fedd$ is typically well below \simba's jet feedback threshold of 0.02, so jet feedback is prevalent.  Hence the low Eddington ratios are strongly correlated with quenched galaxies.  In these systems, there is little cold gas, which means that torque-limited accretion becomes small.  Meanwhile, hot gas is prevalent, making Bondi accretion efficient.  In Angl\'es-Alcaz\'ar et al. (in preparation) we will examine these growth modes in more detail, but here we already see that Bondi-dominated accretion results in significantly more variability in the accretion rate, and hence $\fedd$.  In this regime, there is no obvious correlation of $\fedd$ with $\mbh$.

\label{sec:bhmf}
\begin{figure*}
\begin{center}
\includegraphics[width=0.9\linewidth]{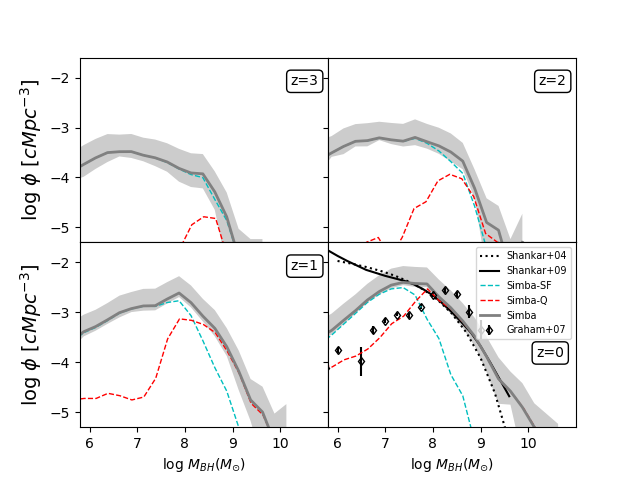}
\captionof{figure}{The evolution of the black hole mass function (BHMF). The grey band shows the BHMF for \simba\ galaxies and the cyan and red lines depict the star forming and quiescent populations within \simba\ respectively. The total BHMF shows good agreement with \citet{Shankar2004} (dotted black line) and \citet{Shankar2009}(solid black line) for $\mbh>10^{7}M_{\odot}$, who estimated the BHMF from the spheroidal mass and velocity dispersion of galaxies, but \simba\ underpredicts the BHMF at lower masses compared to the \citeauthor{Shankar2004} models. The BHMF of \citet{Grahametal2007} (black points with error bars), estimated using the S\'{e}rsic index of ellipticals and bulges, does not agree with the total BHMF produced by \simba, but agrees with the quiescent population at low masses. 
The total number density of black holes decreases with redshift, with the majority of the black hole population becoming dominated by star-forming hosts at higher redshift.}
\label{fig:bhmf}
\end{center}
\end{figure*}

The right panel of Figure \ref{fig:feddtot} shows the evolution of the median Eddington ratios as a function of \textcolor{black}{$\mbh$.}  The obvious trend is that, overall, $\fedd$ is higher at higher redshifts.  This arises from the higher gas fractions and surface densities at high redshift \citep{DAA2015}.  Observations generally suggest that the Eddington ratios of accreting black holes increase with redshift \citep{Kauffmann2009,Lusso2012,Aird2012}, qualitatively consistent with these predictions.

Examining the evolution more carefully, one can see that the anti-correlation slope flattens with increasing redshift, particularly from $z\sim 1\to 4$. At low masses, black holes are always in the radiative mode~\citep[as observed; e.g.][]{Hale2018}, while at higher masses, the Eddington ratios drop more quickly. Black holes with $\mbh\sim 10^9 M_\odot$ are already in place at $z\sim 4$, but they have accretion rates of $5-10\%$ Eddington, whereas at $z=1$ they have $\fedd\ll 1$\%.  From $z=1\to 0$, the growth of hot gas in high-mass halos results in Bondi accretion starting to become dominant for those black holes, and the anti-correlation between $\fedd$ and $\mbh$ is less clear.

In summary, the Eddington ratio predicted in \simba\ drops with both black hole mass and cosmic time.  The $\fedd$ criterion for jet feedback that quenches galaxies thus kicks in for very high mass black holes at high redshift, and the black hole mass scale for jets (and thus quenching) drops with time. Inasmuch as black hole mass is correlated with stellar mass and thus halo mass, this suggests that the halo mass scale where quenching occurs should be higher at high redshift, which agrees with \citet{Hale2018} who found that the halo masses of efficient accretors flattens at high redshift. This is broadly consistent with expectations from the data-constrained equilibrium analytic galaxy formation model of \citet{Mitra2015}, as well as empirical galaxy formation modeling such as \citet{Moster2018}. In this way, the dropping efficiency of torque-limited accretion at both high masses and low redshifts, along with Bondi accretion from the hot gas, helps to enact and maintain quenching in \simba\ galaxies.

\subsection{Distribution Functions}

We have seen that \simba's black hole--galaxy scaling relations as a function of $M_\star$, SFR, or $\mbh$ broadly agree with observations, albeit with some potentially interesting discrepancies. Here we examine the number densities of black holes of a given mass $\mbh$ and Eddington ratio $\fedd$.  Observationally, these can be challenging to determine owing to completeness issues, nevertheless some general trends are evident to which we can compare \simba\ predictions.

\subsubsection{Black Hole Mass Function}

We first consider the black hole mass function (BHMF). Observational estimates of the BHMF usually involve employing correlations of $\mbh$ with global galaxy properties. 
\citet{Shankar2009} used a compilation of X-ray and optical data to determine the AGN luminosity function and model the average growth rate of black holes, making predictions for the local BHMF assuming a single radiative efficiency and Eddington ratio for all black holes, which they compared to observational determinations of the local BHMF based on the $M_{\rm BH}-M_\star$ and $M_{\rm BH}-\sigma$ relations.
\citet{Grahametal2007} used the measured S\'{e}rsic indices of $\sim 10^4$ galaxies from the Millennium Galaxy Catalogue to estimate the BHMF based on the empirical relation between $M_{\rm BH}$ and S\'ersic index from \citet{GrahamDriver2007}. 
We do not attempt to mimic these criteria in detail, owing principally to the fact that this requires structural decomposition of \simba\ galaxies which could potentially be compromised by resolution effects.  Instead, we assume the observations are properly characterising the black hole masses, and compare to these directly.

Figure \ref{fig:bhmf} shows the BHMF predicted by \simba, at $z=3,2,1,0$ (upper left to lower right).  The solid grey line shows the mass function, and the grey band shows the $1\sigma$ uncertainty determined via jackknife subsampling among the 8 simulation sub-octants. At $z=0$, we compare to observations by \citet{Shankar2004} (black dotted line), \citet{Shankar2009} (solid black line) and \citet{Grahametal2007} (black circles with errorbars).  Finally, we subdivide the black hole population into star-forming and quenched populations above and below \textcolor{black}{ sSFR$_{\rm lim}=10^{(-1.8+0.3z)}$~Gyr$^{-1}$}~\citep[as in][]{Dave2019}, shown as the cyan (\simba-SF) and red (\simba-Q) dashed lines, respectively.

Overall, \simba\ produces a black hole mass function that is in very good agreement with observations for $\mbh\ga 10^{7.5-8}M_\odot$.  There is some variance among the different observational determinations, but these are generally within the $1\sigma$ range of \simba\ predictions.  \simba\ produces a turnover at low black hole masses, which is intermediate between the lack of turnover in the \citet{Shankar2004,Shankar2009} determinations, and the \citet{Grahametal2007} measurements.  In \simba, we get this turnover because we seed black holes at $10^4 M_\odot$, and the torque-limited accretion model grows them very quickly until they join the $\mbh-M_\star$ relation, which results in a small number of rapidly-growing small black holes.  If the numbers of these black holes are under-predicted, it could be that \simba's somewhat arbitrary initial seeding and resulting rapid growth phase may not be fully representative of true black hole growth, which would be unsurprising.  Once black holes grow sufficiently large and are stably evolving upwards on the $\mbh-M_\star$ relation, it appears \simba\ produces a $z\approx 0$ black hole population that is in good agreement with data.

The evolution of the BHMF in \simba\ is such that, in general, it decreases towards higher redshift.  In detail, the lowest mass black holes always have a similar number density, owing to seeding at a given $M_\star\approx 10^{9.5}M_\odot$ whose number density also does not evolve much with redshift~\citep{Dave2019}. As time evolves, the black hole population builds up towards a peak at $\mbh\sim 10^8M_\odot$, above which efficient torque-limited accretion becomes less efficient owing to quenching and the diminution of cold gas, resulting in a dropping BHMF above this mass.  The peak in the BHMF at $\mbh\approx 10^{7.5-8}M_\odot$ is therefore a consequence of the interplay between black hole accretion modes and galaxy quenching.  As can be seen, current observational determinations even at $z=0$ are not conclusive on whether such a peak exists, so this represents a reasonably generic prediction of the currently implemented \simba\ accretion model.

The drop in the BHMF to higher redshifts is broadly consistent with observational findings.  \citet{Kelly2012} studied the evolution of the BHMF and compare several models including those from \citet{Shankar2009,Cao2010,Merloni2008} and find that, with redshift, the mass function amplitude decreases.  \citet{RG2016} find that with an increase in redshift, the black hole mass function decreases in amplitude as well as width, similar to \simba\ predictions. They also find that between $z=1$ and $z=0$ the change in amplitude of the black hole mass function is much less rapid than at higher redshifts; this is also consistent with our predictions.  A more detailed quantitative comparisons would involve properly accounting for measurement techniques and selection effects which we leave for future work, but it appears that the broad characteristics of observed BHMF evolution are reproduced in \simba.

Finally, we consider the BHMF split into star forming and quenched galaxies.  The local BHMF is dominated by quiescent galaxies for $\mbh\ga 10^{7.5}M_{\odot}$, because the black hole itself is responsible for quenching those galaxies via jet feedback.  At higher redshifts, the crossover mass scale grows, so that quenched galaxies dominate for $\mbh\ga 10^{8.5}M_{\odot}$ at $z=2$.  As discussed in \S\ref{sec:edd}, the quenching scale drops with time owing to $\fedd(M_\star)$ dropping with time, combined with \simba's assumption that only lower-$\fedd$ black holes can give rise to jets that are responsible for quenching galaxies.  Smaller black holes, in contrast, tend to live in star-forming galaxies.  Galaxies with smaller black holes such as the Milky Way are thus predicted to be predominantly star-forming by \simba.

In summary, the BHMF predicted in \simba\ agrees well with observations for $\mbh\ga 10^{7.5}M_\odot$.  \simba\ predicts a peak in the BHMF around this mass owing to black holes growing rapidly below this mass but slowly above it; observations are inconclusive in the shape of the BHMF at lower masses.  The BHMF is lower at higher redshift, with a less prominent peak owing to less quenching.  Large black holes tend to live in quenched galaxies, with a crossover at $\mbh\ga 10^{7.5}M_\odot$ below which star-forming galaxies dominate at $z=0$; this crossover moves to higher $\mbh$ at higher redshifts.  These results are broadly in agreement with various observational constraints, showing that \simba\ plausibly grows black holes over time.

\subsubsection{Eddington Ratio Distribution}
\begin{figure*}
\begin{center}
\includegraphics[width=0.75\linewidth]{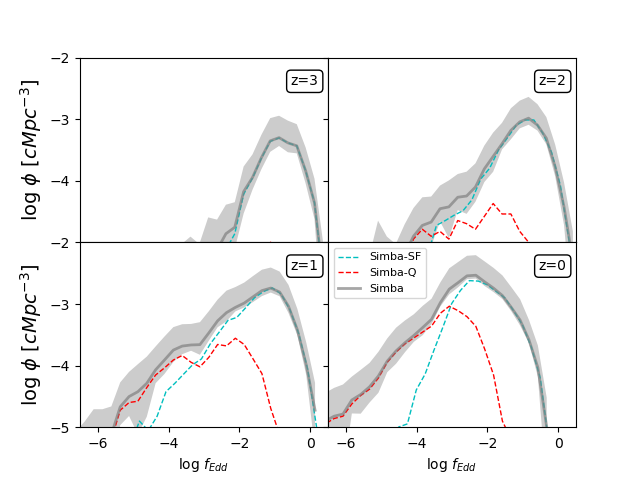}
\captionof{figure}{The evolution of the distribution of Eddington ratios in \simba\ for \textcolor{black}{$z=0-3$}. The grey band shows the total distribution while the cyan and red lines show the star-forming and quiescent populations respectively. High Eddington ratio, or efficiently accreting black holes, dominantly reside in star-forming hosts and by $z\sim2-3$ quiescent galaxies make little to no contribution to the Eddington ratio distribution. The Eddington ratio distribution shifts toward higher values with redshift and show evidence that most black holes are accreting efficiently ($\fedd>1\%$) by $z\sim2-3$.  }
\label{fig:fedddist}
\end{center}
\end{figure*}

Figure \ref{fig:fedddist} shows the distribution of Eddington ratios of all galaxies with black holes in \simba, represented by the grey band (based on jackknife resampling), at $z=3,2,1,0$.  A breakdown into star-forming and quiescent populations is shown by the cyan and red lines, respectively.

The $\fedd$ distribution at every redshift is peaked, with a rapid dropoff to high $\fedd$ and a slower, power-law dropoff to low $\fedd$.  The peak occurs at $\fedd\approx 0.1$ at $z=3$, dropping to $\fedd\approx 10^{-2.5}$ at $z=0$.  This drop in the characteristic Eddington ratio with time was noted in Figure~\ref{fig:feddtot}, and is partly responsible for the increase in jet feedback activity and quenching at later epochs. 
Black holes in star-forming galaxies accrete more efficiently than those in quenched galaxies.  The growing quenched galaxy population creates a tail to low $\fedd$ values that becomes quite prominent at low redshifts, and causes the overall distribution to broaden.  

\begin{figure}
\begin{center}
\includegraphics[width=0.9\linewidth]{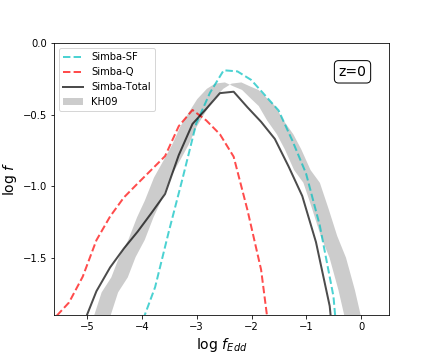}
\captionof{figure}{The fractional distribution of Eddington ratios in \simba\ at $z=0$. The black line shows the total distribution for SMBHs in \simba\ while the red and cyan lines show the fractional distributions from quiescent and star-forming hosts respectively with errorbars removed for clarity. The grey band depicts [O\RomanNumeralCaps{3}] observations from \citet{Kauffmann2009} for SDSS galaxies. The low $\fedd$ end of the distribution is slightly overpredicted by \simba, as well as the high $\fedd$ end is slightly underpredicted. However, the overall distribution of Eddington ratios well trace the observations.} 
\label{fig:fFEDD}
\end{center}
\end{figure}

We compare to observations of the Eddington ratio distribution from
\citet{Kauffmann2009}, who use the [O\RomanNumeralCaps{3}] luminosity of SDSS galaxies to derive $\fedd$ assuming a bolometric correction factor of $\sim600$. With this, $\log$~L[O\RomanNumeralCaps{3}]/$\mbh$ can be converted to $\log\fedd$ by adding $\sim1.7$~dex. Their sample is separated by the amplitude of the 4000\AA~break, $D_{n}4000$, showing that galaxies with $D_{n}4000<1.5$, which are galaxies with recent or ongoing star-formation, follow a lognormal distribution while galaxies with $D_{n}4000>1.8$, which are galaxies with little or no star formation, have a power law distribution. With this, it is deduced that when there is cold gas in the bulge of a galaxy, the central black hole regulates its own growth, and when this cold gas is depleted, the growth of the black hole is regulated by the rate at which evolved stars lose their mass.

Figure \ref{fig:fFEDD} shows the fractional distribution of $\fedd$ in \simba. The total distribution is represented by the black solid line and the red and cyan dashed lines are the quiescent and star-forming fraction distributions respectively split by the same sSFR cut used throughout this paper. The grey band shows observations by \citet{Kauffmann2009} of $D_{n}4000<1.5$ galaxies assuming a range of bolometric corrections $\sim300-600$.  These generally correspond to star-forming systems, so it is more appropriate to compare to the dashed cyan star-forming galaxy predictions from \simba.

For high $\fedd$, the star-forming population in \simba\ is in good agreement with the observations; in \citet{Kauffmann2009}, this regime is dominated by low-$D_{n}4000$ galaxies indicative of star-forming systems.  At low $\fedd$, we see discrepancies in which the $\fedd$ values of star-forming galaxies in \simba\ are somewhat overpredicted. The differences may partly be explained by the fact that we use a sSFR cut and not a $D_{n}4000$ cut to separate the galaxies.  

For quenched galaxies, \simba\ follows a log-normal $\fedd$ distribution shifted to lower $\fedd$ from the star-forming systems.  This is, however, inconsistent with the power-law distribution at low $\fedd$ seen by \citet{Kauffmann2009}.  This may be a consequence of the selection effect since the low $\fedd$ objects tend to be the larger elliptical galaxies that are typically more easily identified in SDSS and thus the low end is typically more biased.  Also, because our black hole accretion model occurs from a kpc-sized region, it is dynamically limited in its ability to capture variability on small timescales.  One might regard the accretion rates in \simba\ as reflective of time-averages over a typical inner disk dynamical time ($\ga 10$~Myr), which would tend to turn an intrinsic power-law distribution in $\fedd$ dominated by short-term variability into a lognormal distribution.

Overall, \simba~ produces a fair agreement with observations of Eddington ratios for star-forming systems, and generally produces lower $\fedd$ values for quenched systems.  Issues with selection effects and variability could be impacting these comparisons, which we will investigate in future work.

\subsection{Black hole dependence on HI}
\label{sec:HI}

\begin{figure*}
\begin{center}
\includegraphics[width=0.9\linewidth]{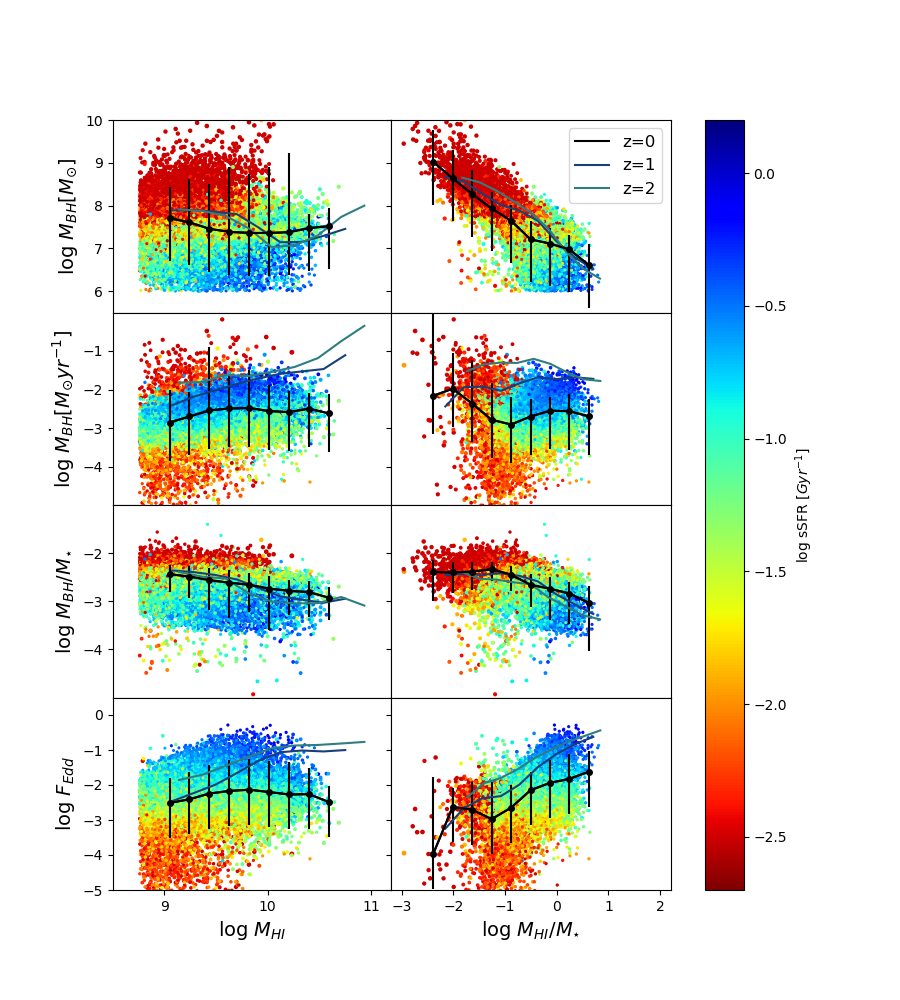}
\vskip-0.3in
\captionof{figure}{Black hole properties as a function of \HI\ content in \simba. Each galaxy at $z=0$ is colour-coded by sSFR and the black line is the running median at $z=0$.  The blue and cyan lines show the corresponding medians at $z=1,2$, respectively. There is a strong anti-correlation between $\mbh$ and $\fHI$, as well as a correlation between $\fedd$ and $\fHI$, both reflecting a tight $\mbh-M_\star$ relation. }
\label{fig:HI}
\end{center}
\end{figure*}

In \simba, the growth phase of black holes is roughly commensurate with that of the stellar content, resulting in a global co-growth of galaxies and their black holes.  This co-growth breaks at late epochs when massive, quenched galaxies appear, whose black holes can grow via Bondi accretion from hot gas that cannot form stars.  Since our star formation model is directly tied to molecular gas content, one expects these trends to also broadly hold for the $H_2$ content of galaxies.  However, the \HI\ cold gas reservoir is not directly tied to star formation, and hence its correlation with black hole mass and accretion is not immediately evident.  Still, the \mufasa\ simulation showed a significant correlation between \HI\ content and SFR~\citep{Dave2017}, and this persists in \simba\ \citep{Dave2019}, which suggests that \HI\ is a reservoir of gas that will ultimately form stars, which should ultimately be also correlated to black hole growth.  The detailed connection between the \HI\ and black hole growth is thus an interesting prediction that connects gaseous fuel in galaxy outskirts with feeding and feedback in the centre of the galaxy.

Owing to present observational challenges, there have been a relatively limited number of studies connecting \HI\ and black holes. \citet{Fabello2011} showed that the \HI\ content of galaxies does not appear to be correlated with black hole accretion in Seyfert galaxies. \citet{HeckmanAndBest} argue that this is expected because the \HI\ lies on larger scales, and is often conspicuously absent in the cores of disk galaxies where the hydrogen is mostly in molecular form.  Hence there is not expected to be an instantaneous connection between \HI\ and black hole growth, just as the instantaneous connection between SFR and $\mdot$ is also weak.  However, in the near future, upcoming radio surveys such as MIGHTEE~\citep{Jarvis2017} and LADUMA~\citep{Holwerda2012} with the new MeerKAT array will soon provide significantly improved data on both \HI-21cm emission as well as black hole accretion than has been available previously.  This will enable larger statistical studies that can identify correlations over longer timescales, and more accurately measure the scatter between gas reservoirs and black hole growth.  It is thus interesting to make predictions for the connection between \HI\ and black hole properties.

Figure \ref{fig:HI} shows how black hole properties depend on the \HI\ content of galaxies in \simba, specifically the total \HI\ mass $M_{\rm HI}$ (left column), and the \HI\ mass fraction $\fHI = M_{\rm HI}/M_{*}$ (right column).  The rows show various black hole properties, from top to bottom: $\mbh$, $\mdot$, $\mbh/M_\star$, and $\fedd$.  All galaxies are colour-coded by sSFR.  The running median at $z=0$ is shown as the black line with $1\sigma$ uncertainties from jackknife resampling.  The blue and turqoise lines show the running medians at $z=1,2$ respectively; the individual galaxy points are not shown at those redshifts.

The top left panel shows that the black hole mass is essentially uncorrelated with \HI\ mass, and there is little evolution in this relation.  More interestingly, at a given \HI\ mass, larger black holes populate more quiescent galaxies.  This shows that there is a strong connection between black hole growth and gas removal in galaxy outskirts, likely owing to suppression of cooling to feed the \HI\ reservoir.
The most star-forming galaxies primarily have the lowest $\mbh$, but there is also a weaker trend that they have the highest $\mHI$.  Hence star formation is enhanced in galaxies that have both small black holes and high gas content.  The top right panel shows $\mbh$ vs. $\fHI$, which displays a strong anti-correlation, reflecting a tight $\mbh-M_\star$ relation at late times, with the most star-forming galaxies having concurrently the smallest black holes and highest gas fractions.  There is modest evolution upwards in this relation with time, such that galaxies at a given black hole mass have higher $\fHI$ at earlier times, reflecting the overall increase in gas content in galaxies at earlier epochs~\citep{Dave2019}.

The second row shows the dependence of BHAR on \HI\ content. At $z=0$, BHAR shows little correlation with \HI\ mass, but there is an evident correlation at $z\ga 1$ such that black hole accretion is stronger for higher \HI\ masses.  At these earlier epochs, the accretion is dominated by the torque-limited mode, which depends on gas fraction.  Even though torque-limited accretion is computed within the core of the galaxy while the \HI\ is more diffusely distributed, the overall enhanced gas content appears to drive black hole accretion.  By $z=0$, in contrast, the emergence of quenched galaxies dominated by Bondi accretion results in no obvious correlation with \HI\ mass.  Even at $z=0$, the star-forming galaxies appear to follow the relations at higher redshift; however, a large population of low-sSFR galaxies overwhelms that trend. Interestingly, at low-$\mHI$, there is a larger scatter of accretion rates in the quenched galaxies, suggesting that Bondi accretion is more stochastic.  For $\fHI$ (right panel), we see that there is no correlation at any redshift with \HI\ fraction, but the most star-forming galaxies have both high $\fHI$ along with higher BHAR.  As discussed previously, the enhanced gas content commensurately drives both stellar and black hole growth.

The third row of Figure~\ref{fig:HI} shows the dependence of $\mbh/M_\star$ on \HI\ properties.  The values are $\mbh/M_\star \approx 10^{-2.5}-10^{-3}$ at every redshift.  In detail, there is a weak anti-correlation, with the highest \HI\ masses having slightly lower $\mbh/M_\star$, independent of redshift. A stronger trend is seen when examining SFR properties, where at a given $\mHI$, galaxies with under-massive black holes are clearly more star-forming.  Quenched galaxies, on the other hand, sit well above the mean relation, with $\mbh/M_\star\sim 10^{-2}$.  This plot most starkly shows that black hole mass is a key governor of whether a galaxy is star-forming or quenched.  The right panel, versus $\fHI$, tells a similar story as the top right panel: galaxies that have the lowest sSFRs have the largest black holes and lowest gas content, and vice versa.  

The bottom row shows the Eddington ratio, which is just a scaled version of $\mdot/\mbh$~(eq.~\ref{eq:fedd}), versus \HI\ properties.  At $z=0$, just like with the BHAR (second row), there is no trend with \HI\ mass, while a trend emerges at higher redshifts. However, unlike for the BHAR, there is a clear trend with $\fHI$.  In terms of $\mHI$, the contours of constant sSFR are essentially horizontal, showing that $\fedd$ is a strong and clear predictor of sSFR at a given \HI\ mass, independent of $\mHI$.  It is interesting that the highest SFR galaxies have both the most undermassive black holes and the highest specific black hole accretion rates, showing that they are in the process of ``catching up" to the typical galaxy in terms of both their black hole and stellar content.  The origin of why some galaxies end up in this state, relative to other galaxies with overmassive black holes that quench more quickly, will be examined in forthcoming work (Cui et al., in preparation).

These trends represent predictions that can be tested against forthcoming large-scale \HI\ surveys, where ancillary data can provide other global galaxy quantities such as the stellar mass, SFR, and black hole mass.  The strong trends relating the black hole mass and accretion rate with SFR at a given \HI\ mass, and as we saw earlier also at a given stellar mass, are direct outcomes of \simba's black hole growth and feedback models. Confirming or falsifying these predictions will be an important test of \simba's black hole evolution model and quenching feedback.

\section{Summary and Discussion}
\label{sec:discussion}

We have presented results from the  $100\hmpc$ \simba\ cosmological hydrodynamic simulation~\citep{Dave2019}. \simba\ employs a novel two-mode subgrid black hole accretion model: gravitational torque-limited accretion~\citep{DAA2017} from cold gas based on the analytic model of \citet{HopkinsQuataert2011}, and Bondi accretion from hot gas as widely used in other galaxy formation simulations.  In this paper we examine the predictions of \simba\ for the growth and evolution of the black hole population relative to their host galaxies, in order to assess the model's broad plausibility and characterise basic predictions of galaxy--black hole co-evolution.  Our main results are as follows:

\begin{itemize}

\item The global black hole mass density and black hole accretion rate density trace the stellar mass and star formation rate densities, respectively. On average, black holes and galaxies grow commensurately, which is broadly consistent with observations.  At $z=0$ in \simba, the ratio of the total $M_\star$ to $\mbh$ density, and also that of SFR to BHAR density, is $\approx 300-400$, which is fairly constant throughout cosmic time.  For stellar to black hole mass density this is mostly in agreement with observations, but for SFR to BHAR density it is an order of magnitude lower than observed, suggesting that current surveys may miss a large fraction of black hole accretion.

\item The mass of black holes is strongly correlated with $M_{*}$, and there is no significant evolution in the $M_{\rm BH}-M_{*}$ relation for $z=0-5$.  There is larger scatter at lower masses, in the regime where black hole seeds are converging onto the $M_{\rm BH}-M_{*}$ relation.  

\item The black hole mass also correlates with stellar velocity dispersion in galaxies, though not quite as tightly as with $M_\star$.  The predictions agree with observational determinations from \citet{KormendyAndHo}. $\mbh-\sigma$ evolves with redshift, but only in a manner that is expected for a universal $M_{\rm BH}-M_{*}$ relation, assuming the expected size evolution of galaxies.

\item The black hole accretion rate $\mdot$ increases with the SFR of its host galaxy for high SFR$\ga 1 M_\odot$yr$^{-1}$, but the relation flattens with significantly more scatter at lower SFR.  The relatively tight correlation for star-forming galaxies arises from a common gas reservoir driving both star formation and black hole growth in the torque-limited mode \citep{DAA2015,DAA2017}.  Bondi accretion dominates for massive black holes in gas poor galaxies, which does not yield a strong correlation with SFR.  The predictions are broadly consistent with observations of BHAR in star forming galaxies, with a hint that it may be slightly higher ($\sim\times 2-3$) than observed.  The predicted $\dot\mbh-SFR$ relation shows no evolution for $z=0-5$.

\item Black hole Eddington rates are strongly anti-correlated with black hole mass at $\mbh\la 10^{8}M_\odot$, with power-law slope nearly $-1$, showing that BHAR is mostly uncorrelated with black hole mass; this is broadly consistent with observations~\citep{Kauffmann2009}.  This is expected in the torque-limited model owing to the weak dependence on $\mbh$ (eq.~\ref{eq:GT2}).  At higher $\mbh\ga 10^8 M_\odot$, Bondi accretion dominates, and the BHAR scales more strongly with black hole mass, resulting in a flatter slope.  The scatter in $\fedd$ becomes very large, indicating that Bondi accretion from hot gas is quite stochastic in \simba.  The $\fedd(\mbh)$ relation evolves fairly strongly towards higher $\fedd$ at a given $\mbh$, with a mild trend for a flattering of the slope at higher $z$.  At $z\ga 2$, the most massive black holes are accreting at reasonably high $\fedd$, broadly consistent with observations of quasars at those epochs.

\item The BHMF predicted by \simba\ shows an increase at the low mass end, and an exponential truncation at the massive end.  This gives a broad peak at $\mbh\approx 10^{7.5-8}M_\odot$, which is where the jet mode feedback in \simba\ kicks in and begins to quench galaxies that also have low $\fedd$.  Splitting galaxies by quenched vs. star-forming clearly shows a dichotomy at this black hole mass scale. The existence of a BHMF peak is thus a direct prediction of \simba's black hole feedback model.  The \simba\ BHMF is in quite good agreement with observations above this peak, but current observational determinations to lower masses are inconclusive about whether there is a peak or not.  We note that the low-mass prediction in \simba\ may owe in part to our seeding prescription, which causes small black holes to grow rapidly; a different seeding prescription may alter the predictions below this peak.  The peak also is less prominent at higher redshifts, because jets are rarer and thus quenching is less effective.  The mass scale at which quenched galaxies dominate also increases in $\mbh$ to higher redshifts, showing that \simba\ predicts that the quenching mass scale downsizes in $\mbh$, and by association, also in $M_\star$ and $M_{\rm halo}$.

\item The Eddington rate function at $z=0$ shows a power-law rise up to $\fedd\approx 10^{-2}$, and then an exponential cutoff above this.  Star-forming galaxies dominate at $\fedd\ga 10^{-3}$, and quenched galaxies only appear at $\fedd\la 0.02$, by which point the jets in \simba\ are ejected at maximum velocity.  The $\fedd$ distribution is in reasonable agreement with SDSS observations from \citet{Kauffmann2009} for star-forming galaxies.  At higher redshifts, the $\fedd$ distribution shifts towards higher values, with a diminishing tail of quenched galaxies resulting in a sharper peak; at $z\ga 2$, $\fedd\ga 0.1$ typically, and we start to see some black holes accreting at the full Eddington rate or even slightly above.

\item The \HI\ content of galaxies shows interesting correlations with black hole mass and accretion rate.  There are at best weak correlations in $\mbh$, $\mdot$, $\mbh/M_\star$, and $\fedd$ with \HI\ mass, but there is always a clear trend that galaxies that are quenched at a given \HI\ mass tend to have large black holes that are accreting inefficiently, while the most star-forming galaxies have undermassive black holes that are accreting efficiently. By examining the \HI\ fraction in galaxies, we see that galaxies are highly star-forming if they have both high gas content and small black holes.  The interplay between gas content, star formation, and black holes is a prediction from \simba\ that can be tested in detail with upcoming multi-wavelength surveys.

\end{itemize}

In general, for where there is observational data, \simba\ reproduces observed black hole--galaxy correlations fairly well, with potential discrepancies such as an
overprediction of BHARs at a given SFR or the underprediction of the BHMF at $M_{\rm BH}<10^{7}M_{\odot}$.  This shows that the new black hole accretion and feedback models in \simba\ are plausible as a platform for studying galaxy--black hole co-evolution and the role of black hole feedback in quenching galaxies.  In upcoming work, we will focus on tracking indvidual black holes to better understand the modes by which black holes grow, examine in more detail how black hole feedback is responsible for quenching, and compare to observational-plane properties such as AGN luminosity functions while more carefully modeling sub-populations of AGN such as high- and low-excitation radio galaxies.  By bringing together results from \simba\ and upcoming surveys of AGN and galaxy evolution, we have a powerful tool to put constraints on the physical mechanisms driving black hole accretion and the extent of its effect on large scale properties of galaxies.

\section*{Acknowledgements}

The authors acknowledge helpful discussions with Philip Best, Weiguang Cui, Katarina Kraljic, \textcolor{black}{ and Francesco Shankar}.  
We thank Philip Hopkins for making \gizmo\ public, and providing our group with early access. We thank Robert Thompson for developing {\sc Caesar}, and the {\sc yt} team for development and support of {\sc yt}.
NT acknowledges support from the South African Radio Astronomy Observatory, which is a facility of the National Research Foundation, an agency of the Department of Science and Technology. NT and RD acknowldedge support from Newton Mobility Grant NMG-R1-180195 from the U.K. Royal Society.
RD acknowledges support from Wolfson Research Merit Award WM160051 from the U.K. Royal Society.
DAA  acknowledges support by a Flatiron  Fellowship. The Flatiron Institute is supported by the Simons Foundation.
The \simba\ simulation was run on the DiRAC@Durham facility managed by the Institute for Computational Cosmology on behalf of the STFC DiRAC HPC Facility. The equipment was funded by BEIS capital funding via STFC capital grants ST/P002293/1, ST/R002371/1 and ST/S002502/1, Durham University and STFC operations grant ST/R000832/1. DiRAC is part of the National e-Infrastructure.

%%%%%%%%%%%%%%%%%%%%%%%%%%%%%%%%%%%%%%%%%%%%%%%%%%

%%%%%%%%%%%%%%%%%%%% REFERENCES %%%%%%%%%%%%%%%%%%

% The best way to enter references is to use BibTeX:

\bibliographystyle{mnras}
\bibliography{BH_SIMBA}

%\bibliography{example} % if your bibtex file is called example.bib

% Alternatively you could enter them by hand, like this:
% This method is tedious and prone to error if you have lots of references
%\begin{thebibliography}{99}
%\bibitem[\protect\citeauthoryear{Author}{2012}]{Author2012}
%Author A.~N., 2013, Journal of Improbable Astronomy, 1, 1
%\bibitem[\protect\citeauthoryear{Others}{2013}]{Others2013}
%Others S., 2012, Journal of Interesting Stuff, 17, 198
%\end{thebibliography}

%%%%%%%%%%%%%%%%%%%%%%%%%%%%%%%%%%%%%%%%%%%%%%%%%%

%%%%%%%%%%%%%%%%% APPENDICES %%%%%%%%%%%%%%%%%%%%%

%\appendix

%\section{Some extra material}

%If you want to present additional material which would interrupt the flow of the main paper,it can be placed in an Appendix which appears after the list of references.

%%%%%%%%%%%%%%%%%%%%%%%%%%%%%%%%%%%%%%%%%%%%%%%%%%

% Don't change these lines
\bsp	% typesetting comment
\label{lastpage}
\end{document}